\documentclass[a4paper,12pt]{article}
\usepackage{amsfonts}
\usepackage{amssymb}
\usepackage{amsmath}
\usepackage{indentfirst}
\usepackage{dsfont}
\usepackage[latin1]{inputenc}

\title{Uniformly accelerating observer in $\kappa$-deformed space-time}
\author{E. Harikumar\footnote{Email: harisp@uohyd.ernet.in}, A. K. Kapoor\footnote{Email: akksp@uohyd.ernet.in} and Ravikant Verma\footnote{Email:ravikant.uohyd@gmail.com}~\\
{\it School of Physics, University of Hyderabad},\\{\it Central University P O, Hyderabad-500046, India}}

\begin{document}

\maketitle

\begin{abstract}
In this paper, we study the effect of $\kappa$-deformation of the space-time on the response function of a uniformly accelerating detector coupled to a scalar field. Starting with $\kappa$-deformed Klein-Gordon theory, which is invariant under a $\kappa$-Poincar\'e algebra and written in commutative space-time, we derive $\kappa$-deformed Wightman functions, valid up to second order in the deformation parameter $a$. Using this, we show that the first non-vanishing correction to the Unruh thermal distribution  is only in the second order in $a$. We also discuss various other possible sources of $a$-dependent corrections to this thermal distribution.
\end{abstract}

\section{Introduction}

Different approaches to quantum theory of gravity suggest that the notion of space-time is modified at the microscopic level and the space-time coordinates 
get quantized. This leads to the modification of the notions of symmetry of the space-time. The symmetry algebra of certain quantum gravity models is known to be the $\kappa$-Poincar\'e algebra and the corresponding space-time is the $\kappa$-Minkowski space-time. Various aspects of $\kappa$-space-time as well as construction and study of field theory models on this space-time are being vigorously investigated in recent times   \cite{luk1,wess,dsr,glikman,mdljlm,sm,us,kappa1,kappa2, us1}. $\kappa$-Poincar\'e algebra and $\kappa$-space-time are also known to be related to the doubly special relativity\cite{dsr}, a modified relativity principle which naturally incorporates a fundamental length scale, required by many approaches to quantum gravity.

In the last couple of years, various authors have studied possible low energy effects of $\kappa$-deformation of the space-time\cite{pos,akk,kumar}. Recently, Unruh effect\cite{unruh,bir,cris} in the non-commutative space-time has been investigated using different approaches\cite{kun,sachin}. It is well known that in the commutative space-time, a system of an uniformly accelerating detector interacting with a massless scalar field in its vacuum is equivalent to an unaccelerated detector which is in the thermal bath of temperature $T=(2\pi k_{B})^{-1}{\alpha}$, where $\alpha$ is the acceleration of the uniformly accelerating detector\cite{unruh,bir,cris}. The modification of this Unruh effect  in the $\kappa$-Minkowski space-time was studied \cite{kun}. In \cite{kun}, authors started with the Klein-Gordon equation in the $\kappa$-space-time and by making a specific choice for the coupling of the detector with the scalar field, analyzed the changes in the Unruh effect. This detector-field interaction term is defined in the $\kappa$-space-time. This necessitated modification of this interaction (compared to the commutative case) so as to guarantee its hermiticity.  It was shown that there are $a=\frac{1}{\kappa}$-dependent contributions to the transition rate  due to the modifications in $G^+(x-x^\prime)\pm G^-(x^\prime-x)$ where $G^\pm$ are the Wightman functions of the scalar theory. In the commutative space-time, the Lorentz invariance implies $G^+(x-x^\prime)- G^-(x^\prime-x)=0$ where as, in $\kappa$-space-time, this is non-vanishing and leads to $a$ dependent correction to the Unruh effect\cite{kun}. As authors pointed out in \cite{kun}, this first order (in $a$) correction is due to the form of the interaction chosen between the detector and the scalar field.

In this paper, we employee an approach in which we start with the $\kappa$-deformed Klein-Gordon theory which is written in the commutative space-time itself (instead of the non-commutative space-time, as considered in \cite{kun}). This scalar theory is invariant under $\kappa$-Poincar\'e algebra\cite{us}. Now that we are dealing with the theory constructed in the commutative space-time, we can use the standard tools of field theory developed for the commutative space-time to analyze this $\kappa$-deformed scalar theory. The usual interaction Lagrangian describing the coupling of the detector to the scalar field, does not get affected by the requirement of hermiticity. Since this model is constructed in the commutative space-time, we can unambiguously define the trajectory of the uniformly accelerated detector, needed for the study of Unruh effect.

In the next section, we summarize the essentials of the $\kappa$-Poincar\'e algebra whose generators involve only operators defined in the commutative space-time. We also present the $\kappa$-deformed Klein-Gordon equation which is invariant under the action of $\kappa$-Poincar\'e algebra. This scalar theory is written in commutative space-time. In section 3, we present the calculation of the propagator of this deformed scalar theory. Using this, in section 4, we calculate the transition rate of the uniformly accelerating detector as the field get excited. This shows explicitly, the $a^2$ dependent modification of Unruh effect due to the $\kappa$-deformation of the space-time. In section 5, we discuss various $a^2$ dependent corrections to the Unruh effect obtained. We also point out possible sources of further $a$ dependent corrections. Our concluding remarks are given in section 6.

\section{$\kappa$-deformed Klein-Gordon theory}

The coordinates of the $\kappa$-Minkowski space-time satisfy a Lie algebra type commutation relation given by
\begin{equation}
[{\hat x}^i, {\hat x}^j]=0, [{\hat x}^0,{\hat x}^i]=ia {\hat x}^i,~ (a=\frac{1}{\kappa}),\label{kappacom}
\end{equation} 
where the deformation parameter $a$ has the dimension of length. It is well known that the symmetry of this space-time is the $\kappa$-Poincar\'e algebra. The defining relations of this algebra explicitly involve the deformation parameter and in the limit $a\to 0$, reduces to the Poincar\'e algebra. Alternatively, one can consider a different realisation of the $\kappa$-Poincar\'e algebra\cite{sm,us} as the symmetry algebra of the above space-time, which we briefly summarize in this section. In this approach taken in \cite{us}, though the defining relations of the algebra are same as that of the usual Poincar\'e algebra, the explicit form of the generators are modified, and these modifications depends on the deformation parameter. In order to construct this  $\kappa$-Poincar\'e algebra, one first demands that the coordinates of the $\kappa$-space-time can be expressed in terms of the commutative coordinates and their derivatives as
\begin{equation}
{\hat x}_\mu=x^\alpha\Phi_{\alpha\mu}(\partial).
\end{equation}
This realization defines a unique mapping between the functions of noncommutative space-time to the functions on commutative space-time.  Imposing further  requirements 
\begin{eqnarray}
\left[\partial_i, {\hat x}_j\right]=\delta_{ij}\varphi(A),\\
\left[\partial_i, {\hat x}_0\right]=ia\partial_i\gamma(A),\\
\left[\partial_0, {\hat x}_i\right]=0, \left[\partial_0, x_0\right]=\eta_{00},
\end{eqnarray}
where $A=-ia\partial_0$, we obtain,
\begin{eqnarray}
{\hat x}_i=x_i\varphi(A),\label{reala}\\
{\hat x}_0=x_0\psi(A)+iax_i \partial_i\gamma(A).\label{realb}
\end{eqnarray}
Using Eqns.(\ref{reala}, \ref{realb}) in Eqn.(\ref{kappacom}), we obtain
\begin{equation}
\frac{\varphi^\prime}{\varphi}\psi=\gamma(A)-1
\end{equation}
where $\varphi^\prime=\frac{d\varphi}{dA}$ satisfying the boundary conditions $\varphi(0)=1, \psi(0)=1, \gamma(0)=\varphi^\prime (0)+1$ and is finite and $\varphi, \psi, \gamma$ are positive functions.

Further demanding that the commutators of the Lorentz  generators with the coordinates  of $\kappa$-deformed space-time must be linear in  ${\hat x}_\mu$ and the generators themselves, and that these commutators should have smooth commutative limit,  lead to just two class of possible realizations. They are parameterized by $\psi=1$ and $\psi=1+2A$. We consider only the former realization here. 

The symmetry of the underlying $\kappa$-space-time is known to be the $\kappa$-Poincar\'e algebra \cite{majid}, which is a Hopf algebra. It was shown in \cite{sm} that one can have an alternate realisation for  the symmetry algebra corresponding to  the $\kappa$-space-time. The generators of this algebra obey\cite{sm}
\begin{equation}
[M_{\mu\nu}, D_\lambda]=\eta_{\nu\lambda}D_\mu-\eta_{\mu\lambda}D_\nu,~~ [D_\mu,D_\nu]=0,\label{udk}
\end{equation}
\begin{equation}
[M_{\mu\nu}, M_{\lambda\rho }]=\eta_{\mu\rho }M_{\nu\lambda } +
\eta_{\nu\lambda}M_{\mu\rho} - \eta_{\nu\rho }M_{\mu\lambda } -   \eta_{\mu\lambda}M_{\nu\rho},\label{diracder1}.
\end{equation}
In the above, we use $\eta_{\mu\nu}={\rm diag}(-1,1,1,1)$.
Note here that the (Dirac) derivatives $D_\mu$  above, transform as  vectors (unlike the usual derivative operators in the $\kappa$-Minkowski space-time). But here the realisation of the generators do have $a$ dependent terms. The explicit form of the Dirac derivatives and $\square$ are 
\begin{eqnarray}
&D_i=\partial_i\frac{e^{-A}}{\varphi},~~
D_0=\partial_0\frac{\sinh A}{A}+ia\nabla^2\left(\frac{e^{-A}}{2\varphi^2}\right),\label{d2}&\\
&\square =\nabla^2\frac{e^{-A}}{\varphi^2}+2\partial_{0}^2 \frac{(1-\cosh A)}{{A^2}} \label{box}&
\end{eqnarray}
where $\nabla^2=\partial_i\partial_i$ and $A=-ia\partial_0$. Note that $\partial_i$ and $\partial_0$ are the derivatives corresponding to the commutative space-time coordinates. The above algebra is also a Hopf algebra, like the $\kappa$-Poincar\'e algebra\cite{sm,us}.

The Casimir of this algebra, $D_\mu D^\mu$,  can be expressed as 
\begin{equation}
D_\mu D_\mu=\square(1-\frac{a^2}{4}\square),
\end{equation}
where the $\square$ operator satisfy 
\begin{equation}
 [M_{\mu\nu},\square]=0,~[\square, {\hat x}_\mu]=2D_\mu.
\end{equation}

It is clear that the Casimir, $D_\mu D^\mu$ reduces to the usual relativistic dispersion relation in limit $a\to 0$. $\varphi$ appearing in the above equations, characterizes arbitrary realizations of the $\kappa$-space-time coordinates in terms of the commutative coordinates and their derivatives\cite{sm}.

Using the Casimir operator on $\kappa$-space-time, generalized Klein-Gordon equation, invariant under the  $\kappa$- Poincar\'e algebra defined in Eqns.(\ref{udk},\ref{diracder1}), is written, as 
\cite{sm,us}
\begin{equation}
\square(1-\frac{a^2}{4}\square)\Phi(x)-m^2\Phi(x)=0\label{kkg}.
\end{equation}
It is clear from the above that the scalar field and the operators appearing in the $\kappa$-deformed Klein-Gordon equation are defined in the commutative space-time itself. The fact that the
generators of the  $\kappa$-Poincar\'e algebra  and the Casimir are expressed in terms of the commutative coordinates and their derivatives is crucial for this. This allows us to use the conventional field theory techniques to study the $\kappa$-deformed Klein-Gordon theory.

The deformed dispersion relation resulting from Eqn.(\ref{kkg}) is
\begin{equation}
\frac{4}{a^2}\sinh^2\left(\frac{ap_0}{2}\right)-p_{i}^2\frac{e^{-ap_0}}{\varphi^2(ap_0)} -\frac{a^2}{4}\left[\frac{4}{a^2}\sinh^2\left(\frac{ap_0}{2}\right)-p_{i}^2\frac{e^{-ap_0}}{\varphi^2(ap_0)}\right]^2=m^2.\label{dis}
\end{equation}
where $p_0=i\partial_0$ and $p_i=-i\partial_i$.

Since the Casimir as well as the $\square$ operator have the same $a\to 0$ limit, the requirement of correct Klein-Gordon equation in the commutative limit does not rule out other possible generalizations,\cite{sm,us}, like,
\begin{equation}
(\square-m^2)\Phi(x)=0.\label{dis1}
\end{equation}

Thus, by re-expressing the noncommutative coordinates in terms of commutative coordinates and their derivatives, the $\kappa$-deformed Klein-Gordon theory (in Eqn.(\ref{kkg}) and Eqn.(\ref{dis1})) is now completely expressed in terms of the commutative field and all operators appearing in the above $\kappa$-deformed Klein-Gordon equation are also defined in the commutative space-time. This allows us to use the well established calculational methods of field theories defined in the commutative space-time. This should be contrasted with \cite{kun} where the starting Klein-Gordon equation is defined on  the noncommutative space-time itself.

\section{$\kappa$-deformed Scalar Propagator}

The Green function corresponding to the massless $\kappa$-deformed Klein-Gordon operator in Eqn.(\ref{kkg}) is
\begin{equation}
G(x-x^\prime)=\int \frac{d^4p}{(2\pi)^4}
\frac{e^{-ip_0(t-t^\prime)+i{\vec p}\cdot({\vec x}-{\vec x}^\prime)}}{(\frac{4}{a^2}\sinh^2(\frac{ap_0}{2})-{\vec p}^2)[1-\frac{a^2}{4}(\frac{4}{a^2}\sinh^2(\frac{ap_0}{2})-{\vec p}^2)]}\label{prop}
\end{equation}
where we have chosen $\varphi(ap_0)=e^{-\frac{ap_0}{2}}$. 
With this choice, the dispersion relation in Eqn.(\ref{dis}) is same as that of the $\kappa$-Poincar\'e algebra in the bi-crossproduct basis\cite{majid}. In deriving the above propagator, we have assumed that the field operators satisfy the standard commutation relations.

The poles of the propagator are 
\begin{eqnarray}
p_0&=&\pm\frac{2}{a}\sinh^{-1}(\frac{ap}{2})+\frac{4\pi in}{a}, n\in {\mathds{Z}}\label{poles},\\
p_0&=&\pm\frac{2}{a}\sinh^{-1}(\sqrt{1+\frac{a^2p^2}{4}})+\frac{4\pi in}{a}, n\in {\mathds{Z}}\label{poles1},
\end{eqnarray}
and they are first order poles. Notice that the periodicity of the Klein-Gordon equation in Eqn.(\ref{dis}) leads to the last terms in the above equations. This leads to infinitely many poles. It is easy to see that in the limit $a\to 0$, we should first set $n=0$ to get the usual Klein-Gordon propagator in the commutative limit. Also, it is interesting to note that the poles in Eqn.(\ref{poles}) are same for the propagator corresponding to the scalar theory described by Eqn.(\ref{dis1}). But the poles in Eqn.(\ref{poles1}) are not shared by the propagator corresponding to the Klein-Gordon equation given in Eqn.(\ref{dis1}).

The positive Wightman function $G^+(x-x^\prime)$ gets the contributions  when $n=0,$ (with positive sign) and when $n<0$ (with $\pm$ sign in Eqns.(\ref{poles},\ref{poles1})). The negative Wightman function  $G^-(x-x^\prime)$ gets the contributions when  $n=0,$ (with negative sign) and when $n>0$ (with $\pm$ sign in Eqns.(\ref{poles},\ref{poles1})).

By direct calculations, it is easy to see that the contributions to $p_0$ integral from the poles in Eqn.(\ref{poles}) with $n< 0$ for $G^+(x-x^\prime)$
and $n>0$ for $G^-(x-x^\prime)$ cancel among themselves and the only contribution is from the pole at $p_0=\pm\frac{2}{a}\sinh^{-1}(\frac{ap}{2})$. Thus the new, infinitely many poles generated due to the periodicity of the $\kappa$-deformed propagator, do not contribute to the Wightman functions. Thus the effect of $\kappa$-deformation comes in only through the modification of the pole with $n=0$ in Eqn.(\ref{poles}).  Thus, we get 
\begin{equation}
G^+(x-x^\prime)=\frac{1}{2(2\pi)^2}\int_{0}^\infty dp\frac{e^{-\frac{2i}{a}\sinh^{-1}(\frac{ap}{2})(t-t^\prime)}}{\sqrt{(1+\frac{a^2p^2}{4})}~|x-x^\prime|}
(e^{-ip|x-x^\prime|}-e^{ip|x-x^\prime|})
\end{equation}
All calculations, up to now are exact as we have kept terms to all orders in the deformation parameter $a$. From now onwards, we keep only terms up to second order in $a$. Thus, from now onwards, we approximate $\frac{2i}{a}\sinh^{-1}(\frac{ap}{2})=ip-\frac{ia^2p^2}{24}$ and $(1+\frac{a^2p^2}{4})^{-\frac{1}{2}}=1-\frac{a^2p^2}{8}$.
With this, we find
\begin{eqnarray}
&G^+(x-x^\prime)= \frac{1}{(2\pi)^2}~\frac{1}{|x-x^\prime|^2-(t-t^\prime)^2}
-\frac{a^2}{4(2\pi)^2} \frac{|x-x^\prime|^2+3(t-t^\prime)^2}{\left[|x-x^\prime|^2-(t-t^\prime)^2\right]^3}&\nonumber\\
&-\frac{a^2}{(2\pi)^2}
\frac{\left[|x-x^\prime|^2+(t-t^\prime)^2\right](t-t^\prime)^2}{\left[|x-x^\prime|^2-(t-t^\prime)^2\right]^4}&\label{G}
\end{eqnarray}
It is of interest to note that the contributions to $G^+(x-x^\prime)$ from the poles in Eqn.(\ref{poles1}) vanishes (up to second order in $a$) and thus the propagators corresponding to the $\kappa$-deformed Klein-Gordon equations in Eqn.(\ref{kkg}) and Eqn.(\ref{dis1}) are identical(up to second order in $a$). 

Note that in the limit $a\to 0$, above Greens function reduces to the correct commutative limit\cite{bir}. The calculation of negative Wightman function, $G^{-}(x-x^\prime)$ also proceeds in the same fashion and it is easy to see that $G^-(x-x^\prime)=G^+(x-x^\prime)$. From Eqn.(\ref{G}), it is clear that, up to second order in $a$, $G^+(x-x^\prime)=G^-(-(x-x^\prime))$. If we include higher order terms in $a$, this may not be true as the exact Greens function should exhibit the loss of Lorentz invariance due to the $\kappa$-deformation. This feature is different from the results of \cite{kun}.These conclusions regarding $G^{\pm}(x-x^\prime)$ would be same for the theory described by Eqn.(\ref{dis1}) also.

\section{ Detector response function}

To analyze how the vacuum of the above theory will be seen by an accelerating observer, we consider a uniformly accelerating detector whose space-time coordinates are given by $x^\mu(\tau)$, where $\tau$ is the proper time of the detector. We consider that the interaction of the detector with the scalar field, up to first order in the deformation parameter $a$, is described by the conventional, hermition,  interaction Lagrangian ${\cal L}_{int}= m(\tau)\phi(x^\mu(\tau))$\cite{unruh,bir}. Notice here that we do not have to modify the interaction to keep the hermiticity as done in \cite{kun}. This is possible here due to that fact that the $\kappa$-deformed scalar theory is expressed in terms of commutative operators and fields. From Eqn.(\ref{kkg}), note that the first non-vanishing $a$ dependent modification to $\kappa$-deformed Klein-Gordon theory is in the second order in $a$(this is true for the theory described by Eqn.(\ref{dis1}) also). This justifies the assumption that the interaction Lagrangian does not receive any modification up to first order in $a$.

As in the commutative space-time, we consider that the field $\phi(x)$ is in the Minkowski vacuum $|0\rangle_M$ and the detector is in its ground state of energy $E_0.$ When the field makes a transition to an excited state, the uniformly accelerating detector also get excited to a state of energy $E>E_0$. One calculates the amplitude of this transition using first order perturbation theory. Here, one also assumes that the time evolution of the uniformly accelerating detector is given by
\begin{equation}
m(\tau)=e^{iH_0\tau}m(0)e^{-iH_0\tau},\label{TEM}
\end{equation}
where $H_0$ is the Hamiltonian describing the detector and $H_0|E\rangle=E|E\rangle$. Thus the transition probability of the detector from the state with energy $E_0$ to get excited to that of energy $E$ is given by
\begin{equation}
|M_{fi}|^2=\sum_{E}|\langle E|m(0)|E_0\rangle |^2{\cal F}(E-E_0){d\tau}\label{amp}
\end{equation}
where the response function is
\begin{equation}
{\cal F}(E)=\int_{-\infty}^{\tau_0} d\tau\int_{-\infty}^{\tau_0}d{\tau^\prime} e^{-iE(\tau-\tau^\prime)} G^+(x(\tau),x(\tau^\prime)).
\end{equation}
Since the $\kappa$-deformed Wightman functions, up to second order in $a$, satisfy $G^+(x-x^\prime)=G^-(-(x-x^\prime))$, only $G^+$ appears in the definition of the response function. This should be contrasted with the result of \cite{kun}. 
From Eqn.(\ref{amp}), we calculate the rate of transition probability,
\begin{equation}
{\cal T}(\tau_0, E)= \sum_{E}|\langle E|m(0)|E_0\rangle|^2\frac{d{\cal F}}{d\tau_0}.\label{trans1}
\end{equation}

To evaluate the transition probability of a uniformly accelerating detector, we express the propagator in terms of the coordinates of the uniformly accelerating detector,i.e.,
$t=\alpha^{-1}\sinh{\alpha \tau}$ and $x=\alpha^{-1}\cosh{\alpha\tau}, y=0=z.$. Using these in the expression of  $G^+$  and Eqn.(\ref{amp}), after lengthy, but straight forward calculations (along the lines of \cite{kun}) we find
\begin{eqnarray}
{\cal T}(\tau_0, {\bar E}, a)&=&\sum_{E}|\langle E|m(0)|E_0\rangle |^2
\left(\frac{1}{2\pi}~ \frac{{\bar E}}{e^{\frac{2\pi {\bar E}}{\alpha}}-1}\right.\nonumber\\
&+&\left.\frac{a^2}{16\pi}~\frac{{\bar E}}{e^{\frac{2\pi {\bar E}}{\alpha}}-1}\left[2 \cosh 2\alpha\tau_0+4 \cosh^2 2\alpha\tau_0+4 \sinh^2(2\alpha\tau_0)\right]\right.
\nonumber\\
&-&\left.\frac{a^2{\bar E}}{8\pi}~\frac{1}{e^{\frac{2\pi {\bar E}}{\alpha}}-1}\left[
\frac{1}{6} (2\cosh 2\alpha\tau_0+1)({\bar E}^2+\alpha^2)-\alpha^2 \cosh 2\alpha\tau_0\right]\right.\nonumber\\
&-&\left.\frac{a^2}{24\pi\alpha^2} {\bar E}({\bar E}^2+\alpha^2)~ \frac{\cosh 2\alpha\tau_0(1+\cosh 2\alpha\tau_0)}{e^{\frac{2\pi {\bar E}}{\alpha}}-1}\right.\nonumber\\ 
&+&\left.\frac{a^2\alpha}{4\pi^2}\sinh 2\alpha\tau_0~{\cal A}+
\alpha \sinh 2\alpha\tau_0 Cosh2\alpha\tau_0~{\cal B}\right.\nonumber\\
&+&\left.\frac{a^2}{8\pi^2\alpha}\sinh 2\alpha\tau_0(1+2\cosh 2\alpha\tau_0){\cal A}\right)\label{trans}
\end{eqnarray}
where ${\bar E}=E-E_0$ and 
\begin{eqnarray}
{\cal A}=\int_{0}^{\infty}\frac{dk}{k}\left(\frac{|{\bar E}-k|^2}{e^{\frac{2\pi |{\bar E}-k|}{\alpha}}-1}-\frac{|{\bar E}+k|^2}{e^{\frac{2\pi |{\bar E}+k|}{\alpha}}-1}\right),\\
{\cal B}= \int_{0}^{\infty}\frac{dk}{k}\left(\frac{1}{e^{\frac{2\pi |{\bar E}-k|}{\alpha}}-1}-\frac{1}{e^{\frac{2\pi |{\bar E}+k|}{\alpha}}-1}\right).
\end{eqnarray}
Eqn.(\ref{trans}) shows the modification to standard thermal distribution in $\kappa$-space-time and we note that allthe correction terms are of the second order in $a$. Note that in the limit $a\to 0$, we get the commutative result.

\section{Discussion}

We have obtained the corrections to the transition rate of the detector from ground state to an excited state as the field makes a transition to an excited state in Eqn.(\ref{trans}). The $a$ dependent terms show the deviation of the Unruh effect  due to $\kappa$-deformation of the space-time. In Eqn.(\ref{trans}), first three $a$ dependent terms have the same Bose-Einstein distribution as in the commutative case, but now multiplied with $a$ dependent weight factors. The remaining terms shows the deviation from the Bose-Einstein distribution due to the $\kappa$-deformation. All the $a$ dependent corrections also have explicit dependence on the detector time $\tau_0$. This shows that different observers (detectors) measure different transition probabilities, showing the loss of Lorentz invariance due to the $\kappa$-deformation. Since the Bose-Einstein distribution is exactly same as in the commutative case, the corresponding temperature does not get any modification due to the $\kappa$-deformation. In Eqn.(\ref{trans}),the $a$ dependent terms do have  explicit dependence on $\tau_0$. Thus as $\tau_0$ changes, the transition rate ${\cal T}(\tau_0, E, a)$ also changes  with a periodicity decided by the periodic functions through which the $\tau_0$ dependence appear in Eqn.(\ref{trans}).

The poles given in Eqn.(\ref{poles1}) do not have smooth commutative limit, but we have seen that they do not contribute to the propagator in Eqn.(\ref{prop})(up to second order in $a$. Thus the propagator of the deformed Klein-Gordon equation has smooth $a\to 0$ limit.

From Eqn.(\ref{poles}), it is clear that the first non-vanishing $a$ dependent correction to the poles contributing to $p_0$ integration are of second order in $a$. This leads to the lowest non-vanishing corrections in the propagator in Eqn.(\ref{G}) to be of second order in $a$ and hence the modification of Unruh effect is in the second order in $a$. But with a different choice of $\varphi$, say, $\varphi=e^{-ap_0}$, it is easy to see that the first non-vanishing corrections to the poles in Eqn.(\ref{poles}) will be in the first order in $a$ itself. With $\varphi=e^{-ap_0}$, we see that the dispersion relation is same as that considered in \cite{kun} and it was shown there that if the coupling of the detector with the scalar field is hermition, the corrections to the first order in $a$ obtained in \cite{kun} will not be present.

Notice that the realization of non-commutative coordinates in terms of commutative ones and their derivatives given in Eqns.(\ref{reala} and \ref{realb}), facilitated the construction of the $\kappa$ Poincar\'e algebra defined in Eqns.(\ref{udk},\ref{diracder1}). Since the generators and Casimir of this algebra were in terms of operators defined in the commutative space-time, the $\kappa$-deformed  Klein-Gordon theory was constructed completely in the commutative space-time\cite{us}. This allowed us to define a hermition coupling between the detector and the scalar field. This explains why there are no  first order (in $a$) corrections to the transition probability (in Eqn.(\ref{trans})).

Another interesting point to note is that the detector was treated as in the commutative space-time. We have seen that the non-relativistic Hamiltonian with the present choice of $\varphi$, viz: $\varphi=e^{-\frac{ap_0}{2}}$ do have a correction that is first order in $a$\cite{akk}. If we include this correction to the Hamiltonian in Eqn.(\ref{TEM}), then the corresponding energy eigenvalues will also get first order $a$ dependent corrections. This will lead to $a$ dependent modification to the Bose-Einstein distribution appearing in the  transition amplitude in Eqn.(\ref{trans}).

Unruh effect can alternatively,  shown by calculating the Bogolubov coefficients relating  the creation and annihilation operators  associated with the quantized field in the left and right Rindler wedges\cite{bir, cris}. Since we have seen that  the $\kappa$-deformed Klein-Gordon theory (with $\varphi=e^{\frac{-ap_0}{2}}$) analyzed here does not have any $a$ dependent correction up to the first order in $a$, the analysis in terms of the Bogolubov coefficients will be same as that in the commutative case(up to order $a$)\cite{cris}. As pointed out above, if we treat the detector in terms of deformed Hamiltonian, we will still get $a$ dependent correction and the energy eigenvalue $E$ appearing in the distribution function would have $a$ dependence. But more importantly, notice that the approach using Bogolubov coefficients, showing that the detector sees scalar particle in a thermal bath, uses the commutation relations between the creation and annihilation operators of the quantized field.
It was shown that, for the $\kappa$-deformed scalar theory described by Eqn.(\ref{dis1}), the associated creation and annihilation operator obey a deformed oscillator algebra\cite{us}, with the choice $\varphi(ap_0)=e^{-\frac{ap_0}{2}}$. This twisted algebra was derived by demanding consistency of the action of the $\kappa$-Poincar\'e algebra, which is a Hopf algebra, and statistics(flip operation)\cite{us}. This deformed oscillator algebra can lead to $a$ dependent changes in the thermal distribution function seen by the detector. These issues will be discussed elsewhere.

\section{Conclusion}

In this paper, we have analyzed the Unruh effect in the $\kappa$-Minkowski space-time, using the recently developed $\kappa$-deformed Klein-Gordon theory\cite{us}. This deformed scalar theory, which is invariant under the $\kappa$-Poincar\'e algebra defined in Eqns.(\ref{udk},\ref{diracder1}), is written in terms of the commutative fields and operators defined in the commutative space-time. This allowed to model the interaction between the field and detector as in the commutative space-time with out any modifications. It was shown that the first non-vanishing corrections to the transition rate is of second order in $a$.

\end{document}